# Response to arXiv:1209.0731 'Erroneous solution of three-dimensional (3D) simple orthorhombic Ising lattices' by Perk

Z.D. Zhang

*Shenyang National Laboratory for Materials Science, Institute of Metal Research and International Centre for Materials Physics, Chinese Academy of Sciences, 72 Wenhua Road, Shenyang, 110016, P.R. China*

Email: zdzhang@imr.ac.cn

N.H. March

*Department of Physics, University of Antwerp, Antwerp, Belgium and Oxford University, Oxford, England*

Email: iris.howard@telenet.be

This paper is a Response to Professor Perk's recent Comment (arXiv:1209.0731). We point out that the singularities of the reduced free energy $\beta f$, the free energy per site f and the free energy F of the 3D Ising model differ at $\beta = 0$. The rigorous proof presented in the Perk's Comment is only for the analyticity of the reduced free energy $\beta f$, which loses its definition at $\beta = 0$. Therefore, all of his objections lose the mathematical basis, which are thoroughly disproved. This means that the series expansions cannot serve as a standard for judging the correctness of the exact solution of the 3D Ising model. Furthermore, we note that there have been no comments on the topology-based approach developed by Zhang for the exact solution of the 3D Ising model.

This paper is a Response to Professor J.H.H. Perk's recent Comment [1] on the exact solution of the three-dimensional (3D) Ising model, derived by one of us (ZDZ) based on two conjectures [2], and our recent paper on conformal invariance in the 3D Ising model [3] (and other earlier published papers). Firstly, we would like to point out that it is hard for us to find anything new in this new Comment, comparing with his three-years-old Comment/Rejoinder (and arXiv posting) [4,5]. He repeats his objections published in [4,5], though this time they have been formulated in more mathematical forms. We think that in this paper, it is unnecessary to repeat all of Zhang's responses in [6,7], and to refute point-to-point Perk's objections to follow-up papers by Zhang and March.

One of the main objections Perk repeated is that the solution obtained in [2] was wrong, because it was based on an incorrect application of the Jordan–Wigner transformation. However, as Zhang responded in [7], this error in the application of the Jordan-Wigner transformation does not affect the validity of the putative exact solution, since the solution is not derived directly from it. The Conjecture 1, i.e., the approach proposed in [2] for dealing with the topological problems of the 3D Ising model by a rotation in an additional dimension, can be applied directly to the corrected formulation after the correct application of the Jordan–Wigner transformation. We emphasize here that the Conjecture 1 serves for the topological problems existing in the 3D Ising model (not for the incorrect formula published in [2]). There have been no comments on this topology-based approach underlying the derivation yet.

Other main objections in Perk's recent Comment [1] and his previously published Comment/Rejoinder [4,5] are limited to the outcome of the calculations in [2]. These objections are based on a misjudgment that the exact solution of the 3D Ising model must pass the series test, and the solution found in [2] contradict so-called exactly known series expansion results. Such argument is based on a belief that there are rigorously established theorems for the convergence of the high-temperature series. However, as pointed out already in [6,7], all the well-known theorems for the convergence of the high-temperature series are rigorously proved only for $\beta$ (= $1/k_B T$) > 0, not for infinite temperature ($\beta = 0$). Exactly infinite temperature has been never touched in those theorems cited in [1,4,5,8] for the free energy per site f of general lattice models with general interactions, since there is a possibility of the existence of a singularity at $\beta = 0$. For instance, Lebowitz and Penrose indicated clearly in p. 102 of their paper [9] that there is no general reason to expect a series expansion of $p$ or $n$ in powers of $\beta$ to converge, since $\beta = 0$ lies at the boundary of the region E of ($\beta$, z) space. Their proof includes $\beta = 0$ only for hard-core potential in section II of their paper [9], not for the Ising model discussed in other sections. Lebowitz and Penrose at the end of the section II used a word of 'implies' as referred to Gallavotti et al.'s work [10]. However, although Gallavotti et al. proved that the radius of convergence is greater than zero, but once again their proof does not touch $\beta$ = 0 [10], since the inequality just above (1), i.e., $\sum_{\substack{T \cap X - \phi \\ T \neq \phi}} \left| K_{\beta \phi'}(X,T) \right| \leq \left[ \exp\left( e^{\beta \|\phi\|} - 1 \right) - 1 \right]$, is invalid for $\beta = 0$.

Although it is evident that the general Theorems above fail in a rigorous proof for the convergence of the high-temperature series at β = 0, Perk still insists to carry out a proof restricted to the 3D Ising model on a simple cubic lattice [1]. In [1], he made an effort to prove rigorously Theorem 2.9, which is then used as a sword to disprove the exact solution of [2]. Unfortunately, this sword is made of wax! Perk claimed that Theorem 2.9 proves rigorously the analyticity of the reduced free energy βf in terms of β at β = 0. However, as will be described in details below, the reduced free energy βf loses its definition at β = 0, and furthermore, it has different behavior with the free energy per site f. Actually, some mathematical tricks have been performed carefully in his procedure, in order to cover the truth that the analyticity of the free energy per site f in terms of β at β = 0 cannot be proved rigorously. Such tricks first appear in Definition 1.4, which defines the free energy per site $f_N$ and its infinite system limit f by eqn (6), but in form of $-\beta f_N$. Then Lamma 2.5 goes on perpetrating the fraud to discuss the singularity of $\beta f_N$, and finally to prove Theorem 2.9 'rigorously' for βf.

In what follows, we will discuss in detail the singularities of the free energy at/near infinite temperature (see also [7] and its arXiv posting 0812.0194). The key issue here is that the behaviours of the reduced free energy βf, the free energy per site f and the free energy F differ at β = 0. Furthermore, both the reduced free energy βf and the free energy per site f lose their definitions at β = 0 so that one has to face directly the behaviour of the free energy F at β = 0.

Let us start from the initial point of the problem to discuss the origin of the singularities at/near infinite temperature. The total free energy of the system is: $F = U - TS = -k_B T \ln Z$. The singularities in the free energy and other thermodynamic consequences (such as the entropy, the internal energy, the specific heat, the spontaneous magnetization, etc) originate from the singularities of the partition function Z. This is why Yang and Lee discussed the phase transition by evaluating the distribution of roots of the grand partition function (i.e., Z = 0) in their general theory [11,12]. In order to describe infinite systems, one usually normalizes the extensive variables that are homogeneous of degree one in the volume, by the volume **V** (or the number of particles N), keeps the density (i. e. the number of particles per volume) fixed and takes the limit for V (or N) tending to infinity. In this sense, one usually defines the thermodynamic limit (N → ∞) for the free energy per site f  by $f = F/N = -k_B T \ln \lambda$ with $\lambda = Z^{1/N}$. By such a procedure, it is expected that one can establish the fact that f converges uniformly to its common limit as N → ∞, namely, it is performed with an assumption (or expectation) that f is finite [11-14]. In this way, one can easily avoid to deal with the total free energy $F = -Nk_B T \ln \lambda$ of the system, which shows singularities at any temperature as N → ∞ and if $\ln \lambda$ is finite. However, at infinite temperature (T = ∞), there still exists a singularity in the free energy per site f that is equal to negative infinite in the case that $\ln \lambda$ is positive and finite. Using the value of the 3D Ising model λ = 2, one also finds that $f = -k_B T \ln 2 = -k_B \ln 2^T$ has a singularity at T = ∞. This is inconsistent with the assumption for the definition of the free energy per site f, and therefore, such definition loses physical significance at T = ∞. This fact also indicates clearly that β = 0 is a special point, differing with other temperatures. It is clear that one has to face directly the total free energy F to study the singularities of the system at T = ∞.

The total free energy of the system can also be written as $F = k_B T \ln Z^{-1}$. Therefore, besides the roots of the partition function Z, one should also discuss the roots of $Z^{-1}$. Writing $z \equiv \exp(-2\beta H)$ and keeping $\beta H$ fixed in the limit $\beta \to 0$, the partition function of an arbitrary lattice with N sites for the Ising model becomes $Z = (z^{1/2}+z^{-1/2})^N$ [5]. It is easily seen that $z^{1/2} + z^{-1/2} > 1$ satisfies the condition for the zeros of the reciprocal of the partition function, i.e., $Z^{-1} = (z^{1/2} + z^{-1/2})^{-N}$. So, the infinite-temperature zeros of $Z^{-1}$, i.e., $Z^{-1} \to 0$, occur at $z = 1$ as $N \to \infty$, $Z = 2^N \to \infty$. Or more explicitly speaking, the zeros are located at $\beta = 0$, $z = 1$. This point of view can be supported by the fact that the singularity behavior of the logarithmic function $\ln x$ in the two cases of $x = 0$ and $x = \infty$ correspond to those in the logarithmic function $\ln y$ with $y = 1/x$ in two cases of $y = \infty$ and $y = 0$, respectively. It indicates that both singularities at the two limits of $Z = 0$ and $Z = \infty$ are actually the same, except for a minus sign, and considerable interest should be paid to both of them. In Definition 1.4 of [1], the negative sign was carefully moved to the left-hand-side of eqn (6), to avoid the discussion on zeros of $Z^{-1}$. But, if one would always try to conceal singularities of $\ln Z^{-1}$ by mathematical tricks, one would find similar tricks to remove singularities of $\ln Z$ also to violate the Yang – Lee Theorem [11,12]. The singularities of the free energy F and the free energy per site f at $\beta = 0$ suggests that two different forms could exist for the high-temperature series expansions of the free energy per site f.

Perk argued in [5] that such singularities of the whole system are not of physical significance, which should be removed by using the reduced free energy per site $\beta f$.

Perk now admits in page 10 of [1] that the free energy per site f diverges at $T = \infty$, but he still insists that this does not correspond to a physical singularity, as the combination $\beta f$ is to be used. So, it is important to evaluate whether one can use the reduced free energy $\beta f$ at $T = \infty$. As stated in Perk's Rejoinder (and arXiv posting) [5], the reduced free energy per site $\beta f$ is often rewritten as $\beta f = \phi(\{K_i\}, h) = \phi(\{\beta J_i\}, \beta H)$ with some function $\phi$. But the error in [1,5] is easily seen as follows: One needs to set $\beta = 1$ to reach an equivalent between $\beta f$ and f. Setting $\beta = 1$ equalizes to $T = 1/k_B \neq \infty$. Therefore, the necessary and sufficient condition for using the dimensionless parameters $K_i = \beta J_i$, (i = 1,2,3) and $h = \beta H$ and setting $\beta = 1$ is $\beta \neq 0$. Thus, setting $\beta = 1$ is loss of generality for $\beta = 0$, and the replacements $J_i \to \beta J_i$, $H \to \beta H$ and $f \to \beta f$ are validated only for $\beta \to 0$ (not for $\beta = 0$). Clearly, all discussions in the Perk's Rejoinder [5] and recent Comment [1] for the reduced free energy $\beta f$ are only valid at the limit $\beta \to 0$, but not 'exactly' at infinite temperature ($\beta = 0$). Clearly, the intrinsic characters of singularities of the zero at infinite temperature are quite different from those at finite temperatures, which cannot be disregarded by the usual process of removing the singularity at finite temperatures by using $-\beta f$.

From the Yang-Lee Theorem [11,12] and the findings above, in the 3D Ising model there indeed exist three singularities: 1) $H = 0$, $\beta = \beta_c$; 2) $H = \pm i\infty$, $\beta \to 0$; 3) $H = 0$, $\beta = 0$. The third singularity is usually concealed in literature by setting $Z^{1/N}$ and dividing the total free energy F by N (equally, disregarding the singularity of zeros of $Z^{-1}$). The point of $\beta = 0$ has been avoided during the procedure of rigorous proof of all the previous theorems for the analyticity of the free energy per site f and also for the convergence of the high-temperature series. The difficulty has been bypassed by using the dimensionless parameters $K_i = \beta J_i$, (i = 1,2,3) and $h = \beta H$ and setting $\beta = 1$. We point out here that the third singularity has physical significance: The 3D Ising system experiences a change from a 'non-interaction' state at $\beta = 0$ to an interacting state at $\beta > 0$. This change of the states is similar to a 'switch' turning off/on all the interactions at/near infinite temperature, resulting in the change of the topological structures and the corresponding phase factors [2,6,7]. The topological difference of $\beta = 0$ and $\beta \to 0$ requires the different dimensions (3D and (3+1)D, respectively) for describing the many-body interacting Ising system. These also support that the high-temperature series expansions of the free energy per site f can have two different forms for infinite temperature and finite temperatures, as revealed in [2].

In section 3 of [1], Perk raised some further remarks and objections on other follow-up papers of [2]. We rebut these criticisms briefly as follows:

The transfer matrix **V** for the 3D Ising model consists of two kinds of contributions: those reflecting the local arrangement of spins and others reflecting the non-local behaviour of the knots. Any procedure (like, low- and high-temperature expansions, Monte Carlo method, ε-expansions, renormalization group, etc.), which takes only the local spin configurations into account (without topological contributions), cannot be correct for the 3D Ising model. This is because the global (topological) effect exists in the 3D Ising system so that the flopping of a spin will sensitively affect the alignment of another spin located far from it (even with infinite distance). These approximation methods have close relations with the same shortcomings, and thus they obtain the close results, but all far from the exact solution. If one used the exact solution of [2] as a standard, the difference between the exact solution and these approximation approaches would be the good evaluation of the non-local contributions of the 3D Ising model. Whether the exact solution can predict the unknown terms of the usual high-temperature expansions is not important, since such expansions are valid only at temperatures very close to infinite temperature, where the global effect can be neglected. The important thing here is that the exact solution can predict all the terms of another high-temperature series for all finite temperatures, which take into account the topological contribution of knots (internal factors) in transfer matrixes. On the other hand, the exact solution of the 3D Ising model does not need to fit with the low-temperature series. This is because the low-temperature series diverges, which suggests that it is falsified and the validity of its leading term is also doubted. The lack of information of the global behaviours of the 3D Ising system is the root of such divergence in the falsified well-known low-temperature series.

It is true that after "Wick rotation" to real time the 3D Ising model relates to a (2+1)-dimensional quantum system. But what we proposed in [2,3,6,7,15] is more than "Wick rotation", and we introduced the fourth dimension to deal with the topological problems of the 3D Ising model, which agrees well with the topological theory. The introduction of the fourth dimension is also important for the time average [6,7,15] and for quaternionic Hilbert space with quaternionic geometric phase [2,3]. The quaternionic form developed in [2] for wave functions of the 3D Ising model agrees well with Jordan-von Neumann-Wigner procedure [16] according to [17-20], and relates closely with well-developed theories, for instance, complexified quaternion [21], quaternionic quantum mechanics [22-24], and quaternion and special relativity [25]. The quaternion-based functions developed in ref. [2] for the 3D Ising models can be utilized to study the conformal invariance in dimensions higher than two [3]. The 2D conformal field theory can be generalized to be appropriate for three dimensions, within the framework of the quaternionic coordinates with complex weights. The 3D conformal transformations can be decomposed into three 2D conformal transformations, where the Virasoro algebra still works in 2D, but only for each 2D complex plane of quaternionic coordinates in the complexified quaternionic Hilbert space [3]. Finally, we note that it is difficult to obtain the high accuracy of numerical results by Monte Carlo method and renormalization group, due to the limited capability of computers dealing with the 3D Ising or Heisenberg model with global effects in thermodynamic limit (infinite) systems (with $2^N$ configurations for Ising model as $N \to \infty$; even much more configurations for Heisenberg model). Thus it should be very careful to judge which system is of Ising or Heisenberg-type based on experimental and numerical results. Furthermore, it is well-known that one cannot distinguish the curves with power exponent $\alpha < 0.2$ and logarithmic exponent $\alpha = 0$, within errorbars of experiments and numerical calculations [2].

In summary, we have disproved the Perk's recent Comment [1]. We have shown that both f and βf lose their definitions at $\beta = 0$, but with different consequences: the free energy per site f could have two different forms for the high-temperature series expansions as revealed in [2]; the reduced free energy per site βf can be used only for finite temperatures (β > 0), not for exactly infinite temperature (β = 0). The Perk's objections [1] are based on errors of mixing the concepts T → ∞ and T = ∞ (i.e., β → 0 and β = 0), and βf and f. His rigorous proof in [1] is only for the analyticity of the reduced free energy βf (its validity is held at β → 0), not of the free energy per site f. Therefore, all the objections of Perk's Comment [1] do not stand on solid ground and have been rejected. The series expansions cannot serve as a standard for disproving the exact solution found in [2] of the 3D Ising model. Furthermore, there have been no comments on the topology-based approach developed by Zhang in [2] for the exact solution of the 3D Ising model.

ZDZ appreciates the National Natural Science Foundation of China (under grant number 50831006). NHM thanks Professors D. Lamoen, and C. Van Alsenoy for making possible his continuing affiliation with the University of Antwerp.

**References**

[1] J.H.H. Perk, arXiv:1209.0731, to be published in Bulletin de la Société des Sciences et des Lettres de Łodź Série: Recherches sur les Déformations.

[2] Z.D. Zhang, Phil. Mag. **87**, (2007) 5309. (see also arXiv:0705.1045)

[3] Z. D. Zhang and N. H. March, arXiv:1110.5527, to be published in Bulletin de la Société des Sciences et des Lettres de Łodź Série: Recherches sur les Déformations.

[4] J.H.H. Perk, Phil. Mag. **89**, (2009) 761. (see also arXiv:0811.1802)

[5] J.H.H. Perk, Phil. Mag. **89**, (2009)769. (see also arXiv:0901.2935)

[6] Z.D. Zhang, Phil. Mag. **88**, (2008) 3097. (see also arXiv:0812.2330)

[7] Z.D. Zhang, Phil. Mag. **89**, (2009) 765. (see also arXiv:0812.0194, specially for the singularities at/near infinite temperature)

[8] F.Y. Wu, B.M. McCoy, M.E. Fisher and L. Chayes, Phil. Mag. **88**, (2008) 3093. (see also arXiv:0811.3876)

[9] J.L. Lebowitz and O. Penrose, Commun. Math. Phys. 11 (1968) 99.

[10] G. Gallavotti and S. Miracle-Solé, Commun. Math. Phys. 7 (1968) 274.

[11] C.N. Yang and T.D. Lee, Phys. Rev., **87**, (1952) 404.

[12] T.D. Lee and C.N. Yang, Phys. Rev., **87**, (1952) 410.

[13] K. Huang, Statistical Mechanics, (John Wiley & Sons Inc., New York and London, 1963) pp. 321-326 and Appendix C.

[14] Quantum Statistical Mechanics, edited by An Editorial Group of the Department of Physics of Beijing University, (Beijing University Press, Beijing, 1987), pp. 90-93.

[15] Z.D. Zhang and N.H. March, J. Math. Chem. **49**, (2011) 1283.

[16] P. Jordan, J. von Neumann, E. Wigner, Ann. of Math. **35** (1934) 29.

[17] J. Ławrynowicz, S. Marchiafava and A. Niemczynowicz, Adv. Appl. Clifford Alg. **20**, (2010) 733.

[18] J. Ławrynowicz, S. Marchiafava and M. Nowak-Kępczyk, Trends in Differential Geometry, Complex analysis and Mathemtical Physics, Proceedings of the 9th International Workshop on Complex Structures, Integrability and Vector Fields, Sofia, Bulgaria, 25 - 29 August 2008 (pp 156-166). edited by K. Sekigawa, V. S. Gerdjikov and S. Dimiev (World Scientific, Singapore), DOI 10.1142/9789814277723_0018.

[19] J. Lawrynowicz, M. Nowak-Kepczyk; O. Suzuki, Inter. J. Bifurcation Chaos, **22**, (2012) 1230003.

[20] J. Lawrynowicz, O. Suzuki and A. Niemczynowicz, Adv. Appl. Clifford Alg. **22,** (2012) 757

[21] S. de Leo and W.A. Rodrigues Jr., Inter. J. Theor. Phys., **36**, (1997) 2725.

[22] D. Finkelstein, J.M. Jauch, S. Schiminovich and D. Speiser, J. Math. Phys. **3**, (1962) 207.

[23] S. Marchiafava and J. Rembieliński, J. Math. Phys. **33**, (1992) 171.

[24] S.L. Adler, Quaternion Quantum Mechanics and Quantum Fields, (Oxford University Press, New York and Oxford, 1995).

[25] S. de Leo, J. Math. Phys. **37**, (1996) 2955.

This is a Rejoinder to Professor J.H.H. Perk's recent Comment (open letter) [1] to our Response [2] to his paper entitled "Erroneous solution of three-dimensional (3D) simple orthorhombic Ising lattices" [3].

We emphasized that all the debates [1-9], up to date, have focused only on whether the high-temperature series can serve as a standard for judging a putative exact solution of the 3D Ising model, not on the validity of the topologic approach developed in [10] and the correctness of its consequence (i.e., the conjectured exact solution).

As already pointed out in our previous responses [2,5,8], all the well-known theorems (including Perk's new approach in [3]) for the convergence of the high-temperature series of the Ising model are rigorously proved only for $\beta$ (= $1/k_BT$) > 0, not for infinite temperature ($\beta = 0$). Lebowitz and Penrose indicated clearly in the abstract of [11] that their proof for the analyticity of the free energy per site and the distribution function of the Ising model is for $\beta > 0$. If the Ising model were a special case of a hard core on the lattice, Lebowitz and Penrose would prove its analyticity for $\beta = 0$ immediately after their proof for a hard core potential. One sentence would be enough for such proof. But, certainly, it is not the case. Their proof for the Ising model is related with the Yang–Lee Theorems [12,13] for $\beta > 0$, and for the analytic of the function $\beta p$. Here, we have to inspect the definition of the hard-core model and the Ising model to discern the difference between them, and also to clarify the conditions of their proof for the hard-core model. The hard-core potential is defined by $\varphi(r) = +\infty$ for $r \leq a$, and $\varphi(r) < \infty$ for $r > a$, where a is a positive constant (a > 0)

[11]. The Ising ferromagnet is isomorphic to a lattice gas with an attractive interaction potential with $\varphi(0) = +\infty$, and $\varphi(r) \leq 0$ for $r \neq 0$ [11]. So, $a = 0$ for the Ising lattice (see also the definition (9) in page 411 of [13]). The key distinction between the two models is whether $a$ is zero or a positive constant. Though Lebowitz and Penrose claimed that the hard-core systems are analytic in $\beta$ at $\beta = 0$, actually, their proof concerns $\beta p$ (the series (4) of [11]), not $p$ itself (For equivalence between $\beta p$ and $p$, setting $\beta = 1$ equalizes to $T = 1/k_B \neq \infty$). Regardless of this flaw in their proof, on the other hand, we can show that their claim for the hard-core systems is not appropriate for the Ising model. Equation (20) of [11] is claimed to define an upper bound on the positive values $|z|$ for hard-core potentials. However, setting $a = 0$ for the Ising model, one would have $|z| < \infty$ since the volume $K_v(0) = 0$. Clearly, behaviors at $\beta = 0$ are different for the hard-core model ($a > 0$) and the Ising model ($a = 0$). Nevertheless, the analyticity in $\beta$ of both the hard-core and Ising models is proved for $\beta > 0$, not for $\beta = 0$.

In their proofs for the Ising model, Lebowitz and Penrose also relied on (or referred to) Gallavotti et al.'s proofs [14-16]. But, it is evident that none of Gallavotti et al.'s proofs for the Ising model touches $\beta = 0$ [14-16]. On the second page (p. 275) of ref. [14] for a detailed proof, they put, for convenience, $\beta = 1$. When they defined $Z_\Lambda(\Phi)$ in eq. (5) of ref. [16], they also set $\beta = 1$. This condition of $\beta = 1$ is contradictory with $\beta = 0$, as mentioned above (see [2] for detailed discussion). The inequality $\sum_{\substack{T \cap X - \phi \\ T \neq \phi}} |K_{\beta \phi'}(X,T)| \leq \left[\exp\left(e^{\beta \|\phi'\|} - 1\right) - 1\right]$ of ref. [15] (or Proposition 1 (i.e., eq. (17)) of ref. [14]) is invalid for $\beta = 0$. All the facts indicate that $\beta = 0$ cannot be included within the radius of convergence, since it makes the condition for such proof invalid.

The Yang–Lee theory [12,13] is only focused on the zeroes of the grand partition function Z, since their interest is the phase transition at finite temperature, not infinite temperature. We have shown in [2] that for behaviors at infinite temperature, one has to study both the zeros and the poles of the grand partition function (i.e., $Z = 0$ and $Z^{-1} = 0$). Both the free energy per site f and the reduced free energy βf lose their definitions at $\beta = 0$, and thus either of them could have two different forms at/below infinite temperature for high-temperature series expansions, as revealed in [10]. For the singularities at/near infinite temperature, we do not mean that there is a phase transition like what we have at the critical point. There is a change of topological phase factors corresponding to the change of topological structures of the 3D Ising system as interactions are turned on.

For the procedure developed in [17] for extending 2D conformal field theory to be appropriate for three dimensions, three independent Virasoro algebras with weight factors ($\text{Re}|e^{i\phi_i}|$ could be zero or non-zero values) guarantee that three independent Virasoro algebras can be written within the 3+1 dimensional space (i.e., 3-sphere), and one does not need to introduce a 6-dimensional (2+2+2) space for it.

We do not want to repeat the criticisms in [10] on disadvantages of numerical techniques (such as Monte Carlo methods), and the criticisms in [2] on the failure of Perk's theorem in [3].

In summary, we show that the analyticity in β of both the hard-core and Ising models is proved for $\beta > 0$, not for $\beta = 0$. The high-temperature series cannot serve as a standard for judging a putative exact solution of the 3D Ising model. Furthermore, the free energy per site f and the reduced free energy $\beta f$ lose their definitions at $\beta = 0$, and thus either of them could have two different forms for high-temperature series expansions at/below infinite temperature. Three independent Virasoro algebras for 3D conformal field theory can be written within the 3+1 dimensional space (i.e., 3-sphere) with weight factors.

ZDZ appreciates the National Natural Science Foundation of China (under grant number 50831006). NHM thanks Professors D. Lamoen, and C. Van Alsenoy for making possible his continuing affiliation with the University of Antwerp.

[1] J.H.H. Perk, arXiv:1209.0731v2. Bulletin de la Société des Sciences et des Lettres de Łódź Série: Recherches sur les Déformations. 63 (2013).

[2] Z.D. Zhang and N.H. March, arXiv:1209.3247. Bulletin de la Société des Sciences et des Lettres de Łódź Série: Recherches sur les Déformations. 62 (2012) as part II.

[3] J.H.H. Perk, arXiv:1209.0731v1. Bulletin de la Société des Sciences et des Lettres de Łódź Série: Recherches sur les Déformations. 62 (2012).

[4] F.Y. Wu, B.M. McCoy, M.E. Fisher and L. Chayes, Phil. Mag. **88**, (2008) 3093.

[5] Z.D. Zhang, Phil. Mag. **88,** 3097 (2008).

[6] F.Y. Wu, B.M. McCoy, M.E. Fisher and L. Chayes, Phil. Mag. **88**, (2008) 3103.

[7] J.H.H. Perk, Phil. Mag. **89,** 761 (2009).

[8] Z.D. Zhang, Phil. Mag. **89**, 765 (2009).

[9] J.H.H. Perk, Phil. Mag. **89**, 769 (2009).

[10] Z.D. Zhang, Phil. Mag. **87,** 5309 (2007).

[11] J.L. Lebowitz and O. Penrose, Commun. Math. Phys. **11**, 99 (1968).

[12] C.N. Yang and T.D. Lee, Phys. Rev., **87**, (1952) 404.

[13] T.D. Lee and C.N. Yang, Phys. Rev., **87**, (1952) 410.

[14] G. Gallavotti and S. Miracle-Solé, Commun. Math. Phys. **7**, 274 (1968).

[15] G. Gallavotti, S. Miracle-Solé and D.W. Robinson, Phys. Lett. **25A**, 493 (1967).

[16] G. Gallavotti and S. Miracle-Solé, Commun. Math. Phys. **5**, 317 (1967).

[17] Z.D. Zhang and N.H. March, arXiv:1110.5527, Bulletin de la Société des Sciences et des Lettres de Łódź Série: Recherches sur les Déformations. 62 (2012) as part I.